\documentclass{INTERSPEECH2023}
\usepackage{multirow}
\usepackage{kotex}
\usepackage{hyperref}
\usepackage{hhline}

\interspeechcameraready

\title{UnitSpeech: Speaker-adaptive Speech Synthesis with Untranscribed Data}
\name{Heeseung Kim$^1$, Sungwon Kim$^1$, Jiheum Yeom$^1$, Sungroh Yoon$^{*1,2}$}

\address{
    $^1$ Data Science and AI Lab, ECE, Seoul National University, Seoul 08826, Korea\\
    $^2$ Interdisciplinary Program in AI, Seoul National University, Seoul 08826, Korea
}
\email{\{gmltmd789, ksw0306, quilava1234, sryoon\}@snu.ac.kr}

\begin{document}
\maketitle
{\let\thefootnote\relax\footnotetext{$*$ Corresponding Author}}
\begin{abstract}
We propose UnitSpeech, a speaker-adaptive speech synthesis method that fine-tunes a diffusion-based text-to-speech (TTS) model using minimal untranscribed data. To achieve this, we use the self-supervised unit representation as a pseudo transcript and integrate the unit encoder into the pre-trained TTS model. We train the unit encoder to provide speech content to the diffusion-based decoder and then fine-tune the decoder for speaker adaptation to the reference speaker using a single $<$unit, speech$>$ pair. UnitSpeech performs speech synthesis tasks such as TTS and voice conversion (VC) in a personalized manner without requiring model re-training for each task. UnitSpeech achieves comparable and superior results on personalized TTS and any-to-any VC tasks compared to previous baselines. Our model also shows widespread adaptive performance on real-world data and other tasks that use a unit sequence as input\footnote{{Code: \href{https://github.com/gmltmd789/UnitSpeech}{https://github.com/gmltmd789/UnitSpeech}}}.

\end{abstract}

\section{Introduction}
As text-to-speech (TTS) models have shown significant advances in recent years \cite{shen2018natural, kim2020glow}, there have also been works on adaptive TTS models which generate personalized voices using reference speech of the target speaker \cite{pmlr-v162-casanova22a, cooper2020zero, wu2022adaspeech, chen2021adaspeech, yan2021adaspeech}. Adaptive TTS models mostly use a pre-trained multi-speaker TTS model and utilize methods such as using target speaker embedding \cite{pmlr-v162-casanova22a, cooper2020zero, wu2022adaspeech} or fine-tuning the model with few data \cite{pmlr-v162-casanova22a, chen2021adaspeech, yan2021adaspeech}. While the former allows easier adaptation compared to the latter, it suffers from relatively low speaker similarities.

Most fine-tuning-based approaches require a small amount of target speaker speech data and may also require a transcript paired with the corresponding speech. AdaSpeech 2 \cite{yan2021adaspeech} proposes a pluggable mel-spectrogram encoder (mel encoder) to fine-tune the pre-trained TTS model with untranscribed speech. Since the mel encoder is introduced to replace the text encoder during fine-tuning, AdaSpeech 2 does not require a transcript when fine-tuning the decoder on the target speaker. However, its results are bounded only to adaptive TTS and show limitations such as requiring a relatively large amount of target speaker data due to its deterministic feed-forward decoder.

Recent works on diffusion models \cite{pmlr-v37-sohl-dickstein15, DDPM} show powerful results on text-to-image generation \cite{rombach2022high} and personalization with only a few images \cite{kumari2022customdiffusion, ruiz2022dreambooth}, and such trends are being extended to speech synthesis \cite{DiffWave, Grad-TTS} and adaptive TTS \cite{Grad-StyleSpeech, Guided-TTS2}. Guided-TTS 2 leverages the fine-tuning capability of the diffusion model and the classifier guidance technique to build high-quality adaptive TTS with only a ten-second-long untranscribed speech. However, Guided-TTS 2 requires training of its unconditional generative model, which results in more challenging and time-consuming training compared to typical TTS models. 

In this work, we propose UnitSpeech, which performs personalized speech synthesis by fine-tuning a pre-trained diffusion-based TTS model on a small amount of untranscribed speech. We use the multi-speaker Grad-TTS as the backbone TTS model for speaker adaptation which requires transcribed data for fine-tuning. Likewise AdaSpeech 2, we introduce a new encoder model to provide speech content to the diffusion-based decoder without transcript. While AdaSpeech 2 directly uses mel-spectrogram as the input of the encoder, we use the self-supervised unit representation \cite{hsu2021hubert} which contains speech content disentangled with the speaker identity to better replace the text encoder. The newly introduced encoder, named unit encoder, is trained to condition the speech content into the diffusion-based decoder using the input unit. For speaker adaptation, we fine-tune the pre-trained diffusion model conditioned on the unit encoder output with a $<$unit, speech$>$ pair of the target speaker. 
By customizing the diffusion decoder to the target speaker, UnitSpeech is capable of performing multiple adaptive speech synthesis tasks that receive transcript or unit as input.

We show that UnitSpeech is comparable to or outperforms baseline models on adaptive TTS and any-to-any VC tasks. We further ablate how each factor of UnitSpeech affects the pronunciation and speaker similarity for adaptive speech synthesis. In addition to samples for evaluation, we provide samples for a wide range of scenarios, including various real-word reference data from YouTube and other tasks using units on demo page\footnote{{Demo: \href{https://unitspeech.github.io/}{https://unitspeech.github.io/}}}.

Our contributions are as follows:

\begin{itemize}
\item To the best of our knowledge, this is the first work that introduces unit representation to utilize untranscribed speech for speaker adaptation.
\item We propose a pluggable unit encoder for pre-trained TTS model, enabling fine-tuning using untranscribed speech.
\item We introduce a simple guidance technique to improve pronunciation accuracy in adaptive speech synthesis.
\end{itemize}

\section{Method}
\begin{figure*}[h]
    \centering
    \includegraphics[width=\linewidth]{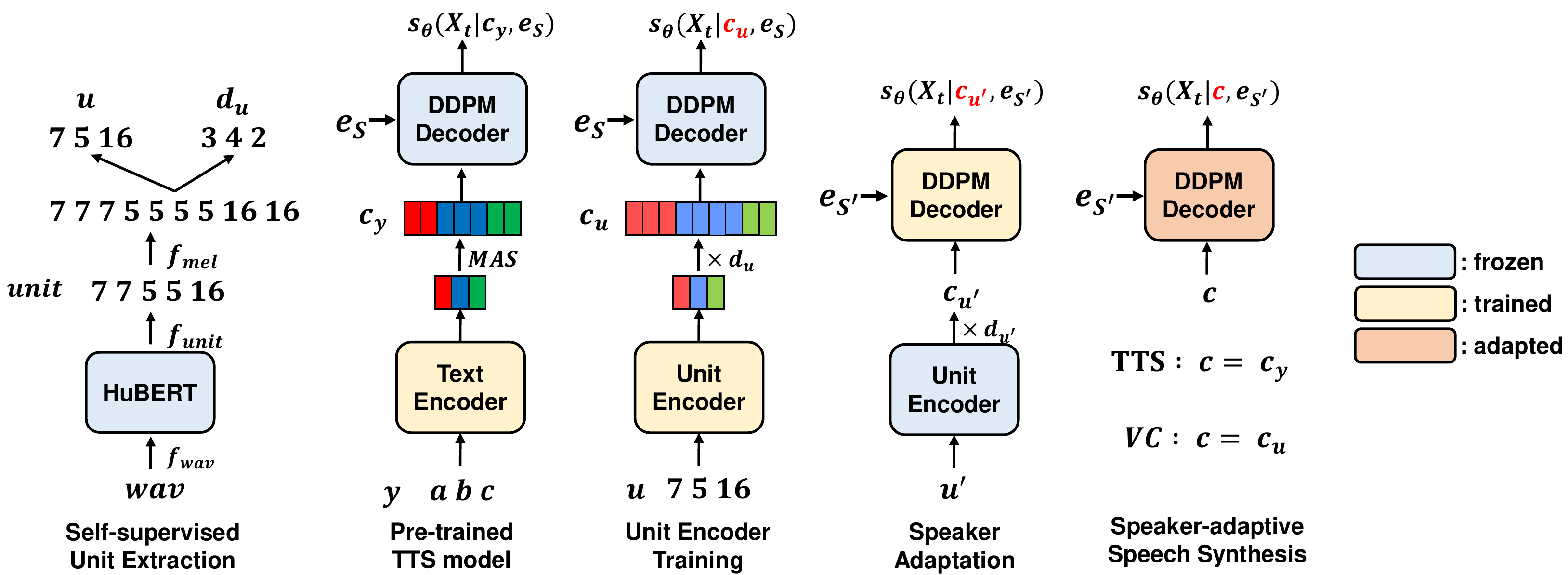}
    \caption{The overall procedure of UnitSpeech.}
    \label{fig1}
    \vskip -0.2in
\end{figure*}

Our aim is the personalization of existing diffusion-based TTS models using only untranscribed data. To personalize a diffusion model \cite{pmlr-v37-sohl-dickstein15, DDPM} without any transcript, we introduce a unit encoder that learns to encode speech content for replacing the text encoder during fine-tuning. We use the trained unit encoder to adapt the pre-trained TTS model to the target speaker on various tasks. We briefly explain the pre-trained TTS model in Section \ref{diffusion}, explain methods used for unit extraction and unit encoder training in Section \ref{unit}, and show how the trained UnitSpeech is used to perform various tasks in Section \ref{speaker-adaptive}.

\subsection{Diffusion-based Text-to-Speech Model}\label{diffusion}
Following the success of Grad-TTS \cite{Grad-TTS} in single-speaker TTS, we adopt a multi-speaker Grad-TTS as our pre-trained diffusion-based TTS model. It consists of a text encoder, a duration predictor, and a diffusion-based decoder, just like Grad-TTS, and we additionally provide speaker information for multi-speaker TTS. To provide speaker information, we use a speaker embedding extracted from a speaker encoder.

The diffusion-based TTS model defines a forward process that gradually transforms mel-spectrogram $X_0$ into Gaussian noise $z=X_T\sim N(0, I)$, and generates data by reversing the forward process. While Grad-TTS defines the prior distribution using mel-spectrogram-aligned text encoder output, we use the standard normal distribution as the prior distribution. The forward process of the diffusion model is as follows: 
\begin{equation}   
    \label{forward diffusion}
        dX_t = -\frac{1}{2}X_t\beta_tdt+\sqrt{\beta_t}dW_t,\quad t\in[0,T],\\
\end{equation}
where the $\beta_t$ is a pre-defined noise schedule, and $W_t$ denotes the Wiener process. We set $T$ to 1 as in \cite{Grad-TTS}.

The pre-trained diffusion-based decoder predicts the score which is required when sampling through the reverse process. For pre-training, the data $X_0$ is corrupted into noisy data $X_t= \sqrt{1-\lambda_t} X_0 + \sqrt{\lambda_t} \epsilon_t$ through the forward process, and the decoder learns to estimate the conditional score given the aligned text encoder output $c_y$ and the speaker embedding $e_S$ with the training objective in Eq. \ref{objective function}.
\begin{equation}   
    \label{objective function}        
        L_{grad}={\mathbb{E}_{t,X_0,\epsilon_t}[\lVert(\sqrt{\lambda_t}s_\theta(X_t,t|c_y,e_S)+\epsilon_t\rVert_2^2]}],
\end{equation}
where $\lambda_t = 1 - {\rm e}^{-\int_0^t\beta_{s}ds}$, and $t\in[0, 1]$. Using the estimated score $s_\theta$, the output of the diffusion-based decoder, the model can generate mel-spectrogram $X_0$ given the transcript and speaker embedding using the discretized reverse process which is as follows: 

\begin{equation}
        \label{discretized reverse diffusion}
        X_{t-\frac{1}{N}} = X_t + \frac{\beta_t}{N}(\frac{1}{2}X_t + s_\theta(X_t,t|c_y, e_S)) + \sqrt{\frac{\beta_t}{N}}z_t,
\end{equation}
where $N$ denotes the number of sampling steps.

In addition to $L_{grad}$ in Eq. \ref{objective function}, the pre-trained TTS model aligns the output of the text encoder with the mel-spectrogram using monotonic alignment search (MAS) proposed in Glow-TTS \cite{kim2020glow} and minimizes the distance between the aligned text encoder output $c_y$ and the mel-spectrogram $X_0$ using the encoder loss $L_{enc}=MSE(c_y, X_0)$. To disentangle the text encoder output with speaker identity, we minimize the distance between the speaker-independent representation $c_y$ and $X_0$ without providing the speaker embedding $e_S$ to the text encoder. 

\subsection{Unit Encoder Training}\label{unit}
While we aim to fine-tune the pre-trained TTS model for high-quality adaptation given minimal amounts of untranscribed reference data, the pre-trained TTS model alone is structurally challenging of doing so. Our pre-trained TTS model is restricted only to training with transcribed speech data, whereas the larger half of real-world speech data is occupied by untranscribed data. As a solution to this problem, we combine a unit encoder with the pre-trained TTS model to expand the generation capabilities for adaptation.

The unit encoder is a model identical to the text encoder of the TTS model in both architecture and role. In contrast to the text encoder which uses transcripts, the unit encoder uses a discretized representation known as unit, which broadens the model's generation capabilities, enabling adaptation on untranscribed speech. Specifically, unit is a discretized representation obtained by HuBERT \cite{hsu2021hubert}, a self-supervised model for speech. The leftmost part of Fig. \ref{fig1} shows the unit extraction process, where speech waveform is used as input of HuBERT, and output representation is discretized by $K$-means clustering into unit clusters, resulting in a unit sequence. Note that by setting an appropriate number of clusters, we can constrain the unit to contain mainly the desired speech content. The obtained unit sequence from HuBERT is upsampled to mel-spectrogram length, where we then compress into unit duration $d_u$ and squeezed unit sequence $u$.

The center of Fig. \ref{fig1} shows the training process of the unit encoder. With squeezed unit sequence $u$ as input, the unit encoder, plugged into the pre-trained TTS model, plays the same role as the text encoder.
The unit encoder is trained with the same training objective $L=L_{grad}+L_{enc}$, only having $c_y$ replaced with $c_u$, an extended unit encoder output using ground-truth duration $d_u$. 
This results in $c_u$ being placed in the same space as $c_y$, enabling our model to replace the text encoder with the unit encoder during fine-tuning.
Note that the diffusion decoder is frozen, and only the unit encoder is to be trained.

\subsection{Speaker-Adaptive Speech Synthesis}\label{speaker-adaptive}
Combining the pre-trained TTS model and the pluggable unit encoder, we are able to perform various speech synthesis tasks in an adaptive fashion by using a single untranscribed speech of the target speaker. Using squeezed unit $u'$ and unit duration $d_{u'}$ extracted from the reference speech as in the previous section, we fine-tune the decoder of the TTS model using the unit encoder. When doing so, the unit encoder is frozen to minimize pronunciation deterioration, and we only train the diffusion decoder using the objective in Eq. \ref{objective function} with $c_y$ switched into $c_{u'}$.

Our trained model is capable of synthesizing adaptive speech using either transcript or unit as input. For TTS, we provide $c_y$ as a condition to the fine-tuned decoder to generate personalized speech with respect to the given transcript. When performing tasks using units including voice conversion or speech-to-speech translation, squeezed unit $u$ and unit duration $d_{u}$ are extracted from the given source speech using HuBERT. The extracted two are inputted into the unit encoder, which outputs $c_{u}$, and the adaptive diffusion decoder uses $c_{u}$ as a condition to generate voice-converted speech.

To further enhance the pronunciation of our model, we leverage a classifier-free guidance method \cite{ho2021classifierfree} during sampling, which amplifies the degree of conditioning for the target condition using an unconditional score. Classifier-free guidance requires a corresponding unconditional embedding $e_\Phi$ to estimate the unconditional score. Since the encoder loss drives the encoder output space close to mel-spectrogram, we set the $e_\Phi$ to the mel-spectrogram mean of the dataset $c_{mel}$ instead of training $e_\Phi$ as in other works \cite{rombach2022high}. The modified score we utilize for classifier-free guidance is as follows:
\begin{equation}   
    \begin{aligned}
    \label{classifier-free guidance}     
        &\hat{s}(X_t, t|c_{c},e_S) = s(X_t, t|c_{c},e_S) + \gamma \cdot \alpha_t, \\ 
        &\alpha_t = s(X_t, t|c_{c},e_S) - s(X_t, t|c_{mel},e_S).
    \end{aligned}
\end{equation}
$c_{c}$ here indicates the aligned output of text or unit encoder while $\gamma$ denotes the gradient scale that determines the amount of provided condition information.

\section{Experiments}
\subsection{Experimental Setup} 
\subsubsection{Datasets} \label{Datasets} We use LibriTTS \cite{zen19_interspeech} to train the multi-speaker TTS model and the unit encoder. LibriTTS is a TTS dataset consisting of 2,456 different speakers, and we use the entire train subset. For training the speaker encoder, we use VoxCeleb 2 \cite{Voxceleb2}, a dataset consisting of 6,112 speakers. To show the unseen speaker adaptation capability of UnitSpeech on TTS, we select 10 speakers and a reference speech for each speaker from the \texttt{test-clean} subset of LibriTTS following YourTTS \cite{pmlr-v162-casanova22a}. For evaluation on any-to-any VC, we randomly choose 10 reference speakers from the \texttt{test-clean} subset of LibriTTS, and randomly select 50 source samples from the \texttt{test-clean} subset. The reference samples are all $7\sim32$ seconds long.

\subsubsection{Training and Fine-tuning Details} 
Our pre-trained TTS model shares the same architecture and hyperparameters with Grad-TTS except for the doubled number of channels for multi-speaker modeling. The architecture of the unit encoder is equal to that of the text encoder. We train the TTS model on 4 NVIDIA RTX 8000 GPUs for 1.4M iterations and train the unit encoder for 200K iterations. We use the Adam optimizer \cite{DBLP:journals/corr/KingmaB14} with the learning rate $1e-4$ and batch size 64. The transcript is converted into the phoneme sequence using \cite{g2pE2019}. When extracting unit sequences, we utilize textless-lib \cite{Kharitonov2022}. We also train the speaker encoder on VoxCeleb2 \cite{Voxceleb2} with GE2E \cite{GE2E} loss to extract the speaker embedding $e_S$ of each reference speech. For fine-tuning, we use Adam optimizer \cite{DBLP:journals/corr/KingmaB14} with learning rate $2\cdot10^{-5}$. We set the number of fine-tuning steps to 500 as a default, which only requires less than a minute on a single NVIDIA RTX 8000 GPU. 

\subsubsection{Evaluation}
To evaluate the performance on adaptive TTS, we compare UnitSpeech with Guided-TTS 2 \cite{Guided-TTS2}, Guided TTS 2 (zero-shot), and YourTTS \cite{pmlr-v162-casanova22a}. For baselines on voice conversion, we used DiffVC \cite{popov2022diffusionbased}, YourTTS \cite{pmlr-v162-casanova22a}, and BNE-PPG-VC \cite{liu2021any}. As for the vocoder, we use the officially released pre-trained model of universal HiFi-GAN \cite{kong2020hifi}. We use the official implementations and pre-trained models for each baseline. Only a single reference speech is used for the adaptation of all the models, and generated audios are downsampled to 16khz for fair comparison. For all the diffusion-based models, we fix the number of sampling steps $N$ to 50. We set the gradient scale $\gamma$ of UnitSpeech to 1.0 for TTS and 1.5 for VC.

We select 5 sentences from \texttt{text-clean} subset of LibriTTS each for the 10 reference speakers chosen in \ref{Datasets} and set the total of 50 sentences as test set for TTS. 50 source speeches for evaluation of VC are selected as explained in \ref{Datasets}. 
We use four metrics for model evaluation: the 5-scale mean opinion score (MOS) on audio quality and naturalness, the character error rate (CER) indicating pronunciation accuracy, the 5-scale speaker similarity mean opinion score (SMOS) and speaker encoder cosine similarity (SECS) to measure how similar the generated sample is to the target speaker.
When calculating CER, we use the CTC-based conformer \cite{gulati20_interspeech} of NEMO toolkit \cite{kuchaiev2019nemo} as Guided-TTS 2. We also use the speaker encoder of Resemblyzer \cite{resemblyzer} for SECS evaluation as YourTTS. We generate adapted samples for each corresponding test sample and measure the CER and SECS values. We report the average values by repeating this measurement 5 times.

\subsection{Results}
\subsubsection{Adaptive Text-to-Speech} 
In Table \ref{tab:main_tts}, we compare UnitSpeech to other adaptive TTS baselines. The MOS results indicate that our model generates high-quality speech comparable to Guided-TTS 2, a model for adaptive TTS only. UnitSpeech also shows superior performance compared to YourTTS, a model capable of both adaptive TTS and voice conversion similar to our model. Furthermore, we show that UnitSpeech is capable of generating speech with accurate pronunciation through the CER results.

We also confirm that our model is on par with Guided-TTS 2, which is also fine-tuned on the reference speech and outperforms zero-shot adaptation baselines on target speaker adaptation from the SMOS and SECS results. Through these results, we show that even though our model is capable of various tasks using either unit or transcript inputs in a personalized manner, it shows reasonably comparable TTS quality against single-task-only baselines. Samples of each model can be found on our demo page.


\begin{table}
\caption{MOS, CER, SMOS, and SECS for TTS experiments on LibriTTS. Guided-TTS 2 (zs) indicates Guided-TTS 2 that performs zero-shot adaptation without fine-tuning.}
\begin{center}
\vskip -0.2in
\begin{tabular}{ccc}
\hline
\multicolumn{1}{|c|}{}                  & \multicolumn{1}{c|}{5-scale MOS}     & \multicolumn{1}{c|}{CER(\%)} \\ \hline\hline
\multicolumn{1}{|c|}{Ground Truth}      & \multicolumn{1}{c|}{$4.49\pm 0.06$}  & \multicolumn{1}{c|}{0.7}     \\ \hline
\multicolumn{1}{|c|}{Mel  + HiFi-GAN \cite{kong2020hifi}}   & \multicolumn{1}{c|}{$4.09\pm 0.10$}  & \multicolumn{1}{c|}{0.75}    \\ \hline
\multicolumn{1}{|c|}{UnitSpeech}        & \multicolumn{1}{c|}{$4.13\pm 0.10$}  & \multicolumn{1}{c|}{1.75}    \\ \hline
\multicolumn{1}{|c|}{Guided-TTS 2 \cite{Guided-TTS2}}      & \multicolumn{1}{c|}{$4.16\pm 0.10$}  & \multicolumn{1}{c|}{0.84}    \\ \hline
\multicolumn{1}{|c|}{Guided-TTS 2 (zs) \cite{Guided-TTS2}} & \multicolumn{1}{c|}{$4.10\pm 0.11$}  & \multicolumn{1}{c|}{0.8}     \\ \hline
\multicolumn{1}{|c|}{YourTTS \cite{pmlr-v162-casanova22a}}           & \multicolumn{1}{c|}{$3.57\pm 0.13$}  & \multicolumn{1}{c|}{2.38}    \\ \hline
\multicolumn{1}{l}{}                    & \multicolumn{1}{l}{}                 & \multicolumn{1}{l}{}         \\ \hline
\multicolumn{1}{|c|}{}                  & \multicolumn{1}{c|}{5-scale SMOS}    & \multicolumn{1}{c|}{SECS}    \\ \hline\hline
\multicolumn{1}{|c|}{Ground Truth}      & \multicolumn{1}{c|}{$3.94 \pm 0.13$} & \multicolumn{1}{c|}{0.933}   \\ \hline
\multicolumn{1}{|c|}{Mel  + HiFi-GAN \cite{kong2020hifi}}   & \multicolumn{1}{c|}{$3.72 \pm 0.13$} & \multicolumn{1}{c|}{0.927}   \\ \hline
\multicolumn{1}{|c|}{UnitSpeech}        & \multicolumn{1}{c|}{$3.90 \pm 0.13$} & \multicolumn{1}{c|}{0.935}   \\ \hline
\multicolumn{1}{|c|}{Guided-TTS 2 \cite{Guided-TTS2}}      & \multicolumn{1}{c|}{$3.90 \pm 0.13$} & \multicolumn{1}{c|}{0.937}   \\ \hline
\multicolumn{1}{|c|}{Guided-TTS 2 (zs) \cite{Guided-TTS2}} & \multicolumn{1}{c|}{$3.71 \pm 0.14$} & \multicolumn{1}{c|}{0.873}   \\ \hline
\multicolumn{1}{|c|}{YourTTS \cite{pmlr-v162-casanova22a}}           & \multicolumn{1}{c|}{$3.34 \pm 0.15$} & \multicolumn{1}{c|}{0.866}   \\ \hline
\end{tabular}
\end{center}
\label{tab:main_tts}
\vskip -0.2in
\end{table}

\subsubsection{Any-to-Any Voice Conversion}
As shown in Table \ref{tab:main_vc}, UnitSpeech performs reasonably on VC task. Our model outperforms baselines regarding naturalness and speaker similarity, with a slight decline in pronunciation accuracy as a trade-off. This result demonstrates that our model is capable of both high-quality adaptive TTS and any-to-any VC. We include samples of our model and baselines on demo page.

\begin{table}
\caption{MOS, CER, SMOS, and SECS for VC experiments on LibriTTS. Mel + HiFi-GAN indicates samples obtained by inputting source speech mel-spectrogram into HiFi-GAN.}
\begin{center}
\vskip -0.2in
\begin{tabular}{ccc}
\hline
\multicolumn{1}{|c|}{}                & \multicolumn{1}{c|}{5-scale MOS}     & \multicolumn{1}{c|}{CER(\%)} \\ \hline\hline
\multicolumn{1}{|c|}{Source}          & \multicolumn{1}{c|}{$4.47\pm 0.06$}  & \multicolumn{1}{c|}{0.7}     \\ \hline
\multicolumn{1}{|c|}{Mel  + HiFi-GAN \cite{kong2020hifi}} & \multicolumn{1}{c|}{$4.24\pm 0.08$}  & \multicolumn{1}{c|}{0.75}    \\ \hline
\multicolumn{1}{|c|}{UnitSpeech}      & \multicolumn{1}{c|}{$4.26\pm 0.09$}  & \multicolumn{1}{c|}{3.55}    \\ \hline
\multicolumn{1}{|c|}{DiffVC \cite{popov2022diffusionbased}}          & \multicolumn{1}{c|}{$3.97\pm 0.09$}  & \multicolumn{1}{c|}{3.67}    \\ \hline
\multicolumn{1}{|c|}{YourTTS \cite{pmlr-v162-casanova22a}}         & \multicolumn{1}{c|}{$3.88\pm 0.10$}  & \multicolumn{1}{c|}{2.20}    \\ \hline
\multicolumn{1}{|c|}{BNE-PPG-VC \cite{liu2021any} }      & \multicolumn{1}{c|}{$3.86\pm 0.10$}  & \multicolumn{1}{c|}{1.37}    \\ \hline
\multicolumn{1}{l}{}                  & \multicolumn{1}{l}{}                 & \multicolumn{1}{l}{}         \\ \hline
\multicolumn{1}{|c|}{}                & \multicolumn{1}{c|}{5-scale SMOS}    & \multicolumn{1}{c|}{SECS}    \\ \hline\hline
\multicolumn{1}{|c|}{Source}          & \multicolumn{1}{c|}{-}               & \multicolumn{1}{c|}{-}       \\ \hline
\multicolumn{1}{|c|}{Mel  + HiFi-GAN \cite{kong2020hifi}} & \multicolumn{1}{c|}{-}               & \multicolumn{1}{c|}{-}       \\ \hline
\multicolumn{1}{|c|}{UnitSpeech}      & \multicolumn{1}{c|}{$3.83 \pm 0.13$} & \multicolumn{1}{c|}{0.923}   \\ \hline
\multicolumn{1}{|c|}{DiffVC \cite{popov2022diffusionbased}}          & \multicolumn{1}{c|}{$3.69 \pm 0.13$} & \multicolumn{1}{c|}{0.909}   \\ \hline
\multicolumn{1}{|c|}{YourTTS \cite{pmlr-v162-casanova22a}}         & \multicolumn{1}{c|}{$3.56 \pm 0.12$} & \multicolumn{1}{c|}{0.763}   \\ \hline
\multicolumn{1}{|c|}{BNE-PPG-VC \cite{liu2021any} }      & \multicolumn{1}{c|}{$3.50 \pm 0.14$} & \multicolumn{1}{c|}{0.851}   \\ \hline
\end{tabular}
\end{center}
\vskip -0.3in
\label{tab:main_vc}
\end{table}


\subsubsection{Other Data and Tasks}
In the previous section, we explained that by fine-tuning the model with a single reference speech of the target speaker, we were able to obtain results either comparable or superior to the baselines on both TTS and VC tasks. UnitSpeech is capable of not only TTS and VC but also any other speech synthesis task that may use unit, providing a sense of personalization to each task. On speech-to-speech translation (S2ST), one of the most general tasks that can utilize unit, we replace the speech synthesis part, which generally uses a single speaker unit-HiFi-GAN \cite{popuri22_interspeech}, with UnitSpeech, and show possibilities of personalized S2ST on CoVoST-2 \cite{wang21s_interspeech}. Samples are on our demo page.

UnitSpeech also maintains reasonable fine-tuning quality even on real-world data for various tasks. To show the real-world availability, we use 10-second-long real-world data extracted from Youtube. Due to copyright issues, we do not explicitly upload these data, but instead, post the Youtube link and start time/end time of each data. We post various adaptation samples on our demo page.

\subsubsection{Analysis}

\begin{table}
\caption{CER, SECS regarding the number of unit clusters, fine-tuning iterations, length of untranscribed speech used for fine-tuning, and the gradient scale in classifier-free guidance.}
\begin{center}
\vskip -0.2in
\label{tab:analysis}
\begin{tabular}{|cccccc|}
\hline
\multicolumn{2}{|c|}{\multirow{2}{*}{}}                                                                                         & \multicolumn{2}{c|}{Text-to-Speech}                        & \multicolumn{2}{c|}{Voice Conversion} \\ \cline{3-6} 
\multicolumn{2}{|c|}{}                                                                                                          & \multicolumn{1}{c|}{CER (\%)} & \multicolumn{1}{c|}{SECS}  & \multicolumn{1}{c|}{CER (\%)} & SECS  \\ \hline\hline
\multicolumn{1}{|c|}{\multirow{4}{*}{\begin{tabular}[c]{@{}c@{}}$K$\\ (\# Units)\end{tabular}}} & \multicolumn{1}{c|}{50}   & \multicolumn{1}{c|}{1.94}     & \multicolumn{1}{c|}{0.932} & \multicolumn{1}{c|}{12.64}    & 0.928 \\ \cline{2-6} 
\multicolumn{1}{|c|}{}                                                                              & \multicolumn{1}{c|}{100}  & \multicolumn{1}{c|}{1.87}     & \multicolumn{1}{c|}{0.930} & \multicolumn{1}{c|}{5.69}     & 0.920 \\ \cline{2-6} 
\multicolumn{1}{|c|}{}                                                                              & \multicolumn{1}{c|}{200}  & \multicolumn{1}{c|}{1.75}     & \multicolumn{1}{c|}{0.935} & \multicolumn{1}{c|}{3.55}     & 0.923 \\ \cline{2-6} 
\multicolumn{1}{|c|}{}                                                                              & \multicolumn{1}{c|}{500}  & \multicolumn{1}{c|}{2.10}     & \multicolumn{1}{c|}{0.932} & \multicolumn{1}{c|}{3.80}     & 0.918 \\ \hline\hline
\multicolumn{1}{|c|}{\multirow{5}{*}{\# Iters}}                                                  & \multicolumn{1}{c|}{0}    & \multicolumn{1}{c|}{1.89}     & \multicolumn{1}{c|}{0.849} & \multicolumn{1}{c|}{3.65}     & 0.845 \\ \cline{2-6} 
\multicolumn{1}{|c|}{}                                                                              & \multicolumn{1}{c|}{50}   & \multicolumn{1}{c|}{2.15}     & \multicolumn{1}{c|}{0.905} & \multicolumn{1}{c|}{3.78}     & 0.893 \\ \cline{2-6} 
\multicolumn{1}{|c|}{}                                                                              & \multicolumn{1}{c|}{200}  & \multicolumn{1}{c|}{1.96}     & \multicolumn{1}{c|}{0.925} & \multicolumn{1}{c|}{3.92}     & 0.924 \\ \cline{2-6} 
\multicolumn{1}{|c|}{}                                                                              & \multicolumn{1}{c|}{500}  & \multicolumn{1}{c|}{1.75}     & \multicolumn{1}{c|}{0.935} & \multicolumn{1}{c|}{3.55}     & 0.923 \\ \cline{2-6} 
\multicolumn{1}{|c|}{}                                                                              & \multicolumn{1}{c|}{2000} & \multicolumn{1}{c|}{2.04}     & \multicolumn{1}{c|}{0.937} & \multicolumn{1}{c|}{3.78}     & 0.925 \\ \hline\hline
\multicolumn{1}{|c|}{\multirow{3}{*}{\begin{tabular}[c]{@{}c@{}}Length\\ (secs)\end{tabular}}}      & \multicolumn{1}{c|}{3}    & \multicolumn{1}{c|}{2.16}     & \multicolumn{1}{c|}{0.916} & \multicolumn{1}{c|}{3.82}     & 0.926 \\ \cline{2-6} 
\multicolumn{1}{|c|}{}                                                                              & \multicolumn{1}{c|}{5}    & \multicolumn{1}{c|}{1.96}     & \multicolumn{1}{c|}{0.921} & \multicolumn{1}{c|}{3.44}     & 0.925 \\ \cline{2-6} 
\multicolumn{1}{|c|}{}                                                                                  & \multicolumn{1}{c|}{30}   & \multicolumn{1}{c|}{1.88}     & \multicolumn{1}{c|}{0.949} & \multicolumn{1}{c|}{3.07}     & 0.946 \\ \hline\hline
\multicolumn{1}{|c|}{\multirow{5}{*}{\begin{tabular}[c]{@{}c@{}}Gradient\\ scale $\gamma$\end{tabular}}}     & \multicolumn{1}{c|}{0.0}  & \multicolumn{1}{c|}{2.83}     & \multicolumn{1}{c|}{0.941} & \multicolumn{1}{c|}{5.02}     & 0.939 \\ \cline{2-6} 
\multicolumn{1}{|c|}{}                                                                              & \multicolumn{1}{c|}{0.5}  & \multicolumn{1}{c|}{2.04}     & \multicolumn{1}{c|}{0.939} & \multicolumn{1}{c|}{4.15}     & 0.936 \\ \cline{2-6} 
\multicolumn{1}{|c|}{}                                                                              & \multicolumn{1}{c|}{1.0}  & \multicolumn{1}{c|}{1.75}     & \multicolumn{1}{c|}{0.935} & \multicolumn{1}{c|}{3.86}     & 0.93  \\ \cline{2-6} 
\multicolumn{1}{|c|}{}                                                                              & \multicolumn{1}{c|}{1.5}  & \multicolumn{1}{c|}{1.74}     & \multicolumn{1}{c|}{0.929} & \multicolumn{1}{c|}{3.55}     & 0.923 \\ \cline{2-6} 
\multicolumn{1}{|c|}{}                                                                              & \multicolumn{1}{c|}{2.0}  & \multicolumn{1}{c|}{1.79}     & \multicolumn{1}{c|}{0.923} & \multicolumn{1}{c|}{3.74}     & 0.918 \\ \hline
\end{tabular}
\end{center}
\vskip -0.3in
\end{table}

We show the effects of several factors of our model in Table \ref{tab:analysis}.

\textbf{The number of unit clusters} We observed that the number of clusters $K$ does not significantly affect TTS results. In the case of voice conversion, however, which directly uses units as inputs, the increase in $K$ allows a more precise segmentation of pronunciation, leading to better pronunciation accuracy.

\textbf{Fine-tuning} 
Our results demonstrate that the more we fine-tune, speaker similarity increases gradually and eventually converges around 500 iterations. We also observe that the pronunciation accuracy decreases when fine-tuning over 2,000 iterations. Thus, we have set the default number of iterations for fine-tuning to 500, which only takes less than a minute in a single NVIDIA RTX 8000 GPU.

We also measure pronunciation accuracy and speaker similarity according to the amount of reference speech used for fine-tuning. Our results show that both metrics improve as the length of reference speech increases. Furthermore, our model can still achieve sufficient pronunciation accuracy and speaker similarity even with a 5-second-long short reference speech.

\textbf{Gradient scale in classifier-free guidance} The results in Table \ref{tab:analysis} indicate that the proposed guidance method improves pronunciation at the cost of a minor decrease in speaker similarity. Therefore, we choose the gradient scale $\gamma$ that maximizes the pronunciation improvement while minimizing the reduction in speaker similarity, which is 1 for TTS and 1.5 for VC.

\section{Conclusion}
We proposed UnitSpeech, a diffusion model that enables various adaptive speech synthesis tasks by fine-tuning a small amount of untranscribed speech. UnitSpeech consists of a unit encoder in addition to the text encoder, eliminating the need for a transcript during fine-tuning. We also introduce a simple guidance technique that allows UnitSpeech to perform high-quality adaptive speech synthesis with accurate pronunciation. We showed that UnitSpeech is on par with the TTS baselines and outperforms VC baselines regarding audio quality and speaker similarity. Our demo results also indicate that UnitSpeech can robustly adapt to untranscribed speech of real-world data and we can substitute UnitSpeech for speech synthesis modules of tasks that take the unit as input.

\section{Acknowledgements}
This work was supported by SNU-Naver Hyperscale AI Center, Samsung Electronics (IO221213-04119-01), Institute of Information \& communications Technology Planning \& Evaluation grant funded by the Korea govern-
ment (MSIT) [2021-0-01343, AI Graduate School Program (SNU)], National Research Foundation of Korea grant funded by MSIT (2022R1A3B1077720), and the BK21 FOUR program of the Education and Research Program for Future ICT Pioneers, SNU in 2023.

\bibliographystyle{IEEEtran}
\bibliography{main}

\end{document}